# Theory of Operation of Direct String Magnetic Gradiometer with Proportional and Integral Feedback

**Alexey V. Veryaskin**

Gravitec Instruments, School of Physics, University of Western Australia, 35 Stirling Highway, Nedlands, Perth WA6009, Australia

**Abstract**

A quantitative theory of operation of a novel device, namely Direct String Magnetic Gradiometer (DSMG), is presented. The paper provides a detailed analysis of DSMG basic functions and measured quantities, represented in terms of physical parameters that are known either a priori or can be experimentally determined. It leaves a reasonable degree of freedom to further investigate some of the finer detail of this new instrument based on experimental results coming both from the laboratory environment and from field trials. The analysis also allows us to quantitatively evaluate the error budget for an optimised DSMG.

## 1. Introduction

Recently a novel magnetic gradiometer has been developed, where a vibrating string, driven by an AC current, is used as a single sensitive element [1-3]. It operates at room temperature and is designed to directly measure off-diagonal components of the magnetic gradient tensor [4], i.e. – Bxz, Byz and Bxy, provided the distance to an object creating magnetic anomalies is much larger than the length of the string [1].

If the external drive frequency is tuned to one of the string's fundamental harmonics, this will set the string to a resonant motion due to the Ampere force per unit length

$$\frac{\partial \mathbf{f}}{\partial z} = i_s [\mathbf{e}_z \times \mathbf{B}] Sin(\omega t) \tag{1}$$

where $\mathbf{e}_z$ is a unity vector along Z direction chosen to point along the string's length, $i_s$ – is the amplitude of the AC drive current, $\omega$ – is the string's drive angular frequency and **B** is the magnetic induction vector. When the string is a stretched thin flat wire [3] the resonant motion is strictly one-dimensional with its sensitivity axis pointing perpendicular to the plane of motion. In this case the string represents a one-dimensional mechanical harmonic oscillator having an infinite number of resonant (violin) modes [1]

$$\frac{d^2}{dt^2} X_n + \frac{2}{\tau} \frac{d}{dt} X_n + \omega_n^2 X_n =$$
$$= \left[ \frac{2}{\pi n} \left(1 - (-1)^n\right) B_y - (-1)^n \frac{2l}{\pi n} B_{yz} \right] \frac{i_s}{\eta} Sin(\omega t) + N_n(t) \tag{2}$$

where n = 1,2,3…, (for definitions of terms see Section 2).

As follows from Eq.(2), the conventional magnetic field term is coupled only to the odd resonant modes, while the magnetic gradient term of the driving force is coupled to both odd resonant modes and exclusively to the even ones. Based on this intrinsic discrimination between a uniform magnetic induction component and its spatial derivative, the magnetic gradiometer has been built and tested both in the laboratory environment and in the field [2-3].

The sensor module contains no metal parts except an aluminium stretched flat wire, clamped at both ends and housed in a Torlon-made solid rectangular frame. Torlon is an industrial grade plastic whose thermal expansion coefficient precisely matches that of aluminium. An alternating current passes through the wire, forcing it to vibrate at the second violin mode (~850 Hz) coupled to the magnetic gradient. The amplitude of the second violin mode is therefore the measure of a magnetic gradient along the wire.

The laboratory and field tests have shown that the DSMG developed to date is capable of measuring magnetic gradients below 0.2 nT/m in an unshielded environment between the 0.001 Hz to 0.3 Hz frequency range, and work is progressing on increasing its sensitivity to at least 0.02 nT/m per 1 sec average.

In this paper we present a complete theory of the magnetic gradient measurements based on the vibrating string concept. We consider the real experimental situation when the string is an integral part of a dynamic feedback loop. The dynamic properties of such a complex system are very different from those of a stand-alone mechanical oscillator as described in [1]. In particular, there are a number of additional parameters that can play a crucial role in creating an optimized DSMG with the possibility of implementing effective noise suppressing algorithms, such as electronic cooling [7].

The experimental results obtained to date show a very good agreement with the theoretical expectations for the particular sensor design [3]. However, the purpose of this work is to develop a theory, which can describe and quantitatively predict results for any further improvements that would provide a better performance of the sensor design and/or the signal processing. The analysis presented below also allows us to quantitatively estimate an error budget for an optimised DSMG. A detailed description of the experimental data obtained from the laboratory and field tests, and a comparison with the theory presented in this paper will be published elsewhere.

**2. DSMG closed loop operation**

It is assumed that the string vibrates in the XOZ plane of its local reference frame, the origin of which coincides with a lower clamp point. The upper clamp point determines the string's length $l$. $X_n(t)$ is the amplitude of an n-mode mechanical displacement of the string from its unperturbed position aligned with the Z-axis. It is also assumed that all non-linear terms can be ignored as, in fact, the maximum possible mechanical displacements do not exceed sub nanometre scale [5].

Also, in Eq.(2), $\eta$ is the string's mass per unit length, and $\tau$ is its mechanical relaxation time, which is the same for all resonant modes of the string. $N_n(t)$ represents the fundamental thermal noise source (in terms of acceleration noise), which sets an absolute limit on the sensitivity of a DSMG. It has the following correlation function in the white noise area [1]

$$\left\langle N_n(t_1) N_m(t_2) \right\rangle = \frac{8 k_B T}{\eta l \tau} \delta_{nm} \delta(t_1 - t_2) \tag{3}$$

where $k_B = 1.4 \times 10^{-23}$ J/K is the Botzmann constant and T is absolute temperature.

Fig.1 shows a block diagram of feedback loop of the whole system "string oscillator + feedback". The string vibrates at its second resonant mode in the presence of an external magnetic gradient Byz. The mechanical displacements of the string are read out with the use of a mechanical-displacement-to-voltage transducer, which also provides a signal gain. In the DSMG developed to date, the transducer is based on a differential L-C resonant tank [6]. The tank is inductively coupled with two pick-up coils placed at one quarter and three quarters positions along the string. The string is pumped with an RF current tuned to the resonant frequency of the tank. Therefore, its displacements toward or away from the pick-up coils create an electromotive force in the tank due to a mutual inductance between the string and the pick-up coils. The amplitude of the RF voltage across the tank depends on the second violin mode displacements of the string provided the coils are connected differentially.

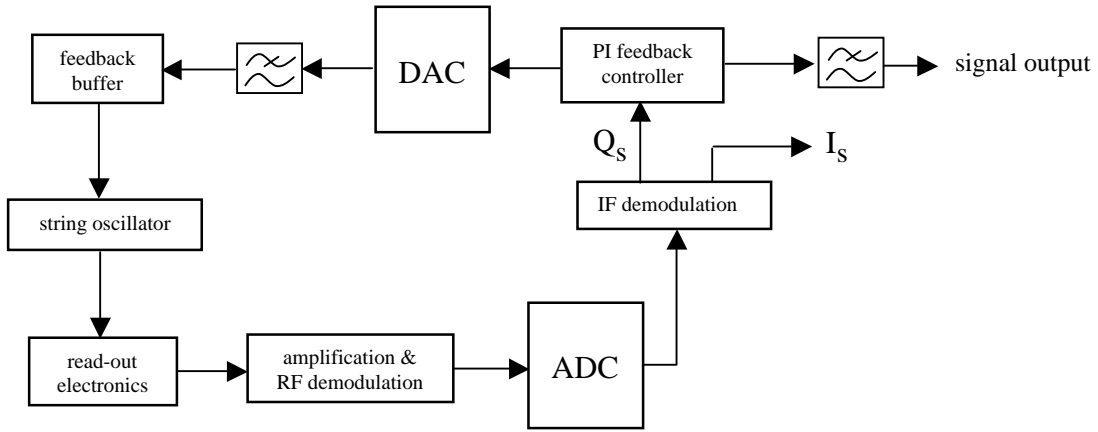

Fig.1. Block-diagram of the feedback loop servo.

In the frequency domain, the RF signal contains the RF carrier frequency and two sidebands formed by the second violin mode vibration frequency. The RF carrier builds up due to a limited common mode rejection factor of the differential pick-up coils (see Section 5 below). In real operation, the RF carrier is suppressed by a reference signal pumped directly into the tank in such a way that it nulls out the RF voltage across the tank (RF carrier null servo is an integral part of the digital signal processing). This allows us to amplify the sidebands further to the point where the signal is RF-demodulated and the signal gain is high enough for signal digitization with a sigma-delta 24-bit ADC.

In the digital domain the signal is mixed with in-phase and quadrature components of the string drive current, i.e. - Sin($\omega$t) and Cos($\omega$t). The quadrature-demodulated component is then used to form the proportional and integral terms of the feedback signal that goes straight back to the string oscillator through an 18-bit DAC and a low pass filter. Its purpose is to maintain a constant value effective magnetic gradient acting upon the string. This is done by the use of a set of feedback coils positioned symmetrically and closed to the string at one quarter and three quarters of the string length. The feedback coils are RF-decoupled from the pick-up coils as they are positioned close to each other in the sensor frame. The feedback signal is converted into a feedback current passing through the feedback coils, which in turn creates a magnetic gradient along the string. Any variation of the magnetic gradient from an external source is cancelled by the feedback magnetic gradient created by the feedback current. The output DSMG data is the integral term of the feedback signal, which passes through a digital output low-pass filter and is monitored and stored by a controlling PC.

In the ideal case when the string drive frequency is exactly tuned to the second resonant mode, the in-phase-demodulated component is absent after the IF demodulation stage. By measuring this component it is possible to determine an error in fine-tuning the string drive frequency, and adjust it accordingly in real time operation. This algorithm forms the string drive frequency loop servo that is shown in Fig.2. It allows us to lock the drive frequency to the second resonant mode of the string within sub mHz accuracy. In fact, a string locked to its second resonant mode and kept at a constant oscillation amplitude is a direct analogy to an IF stage of standard modulation-demodulation based super heterodyne RF receivers.

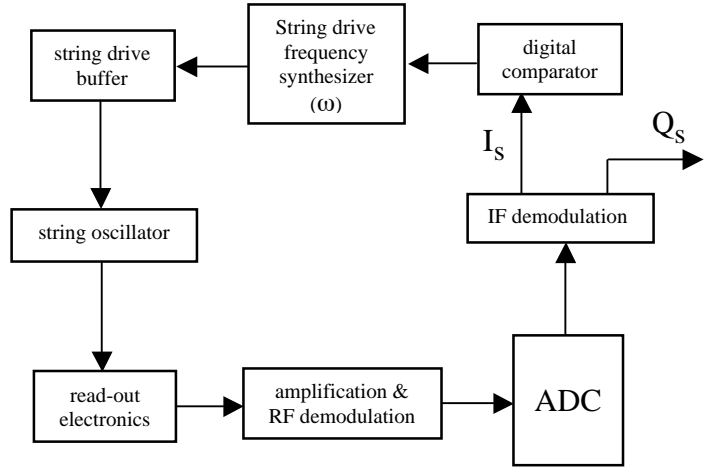

Fig. 2. Block-diagram of the string drive frequency loop servo.

Below we consider the most general case without specifying any particular parts of the system. We introduce the following directly measurable and/or controlled quantitative parameters needed for the system characterization:

$\omega_2$ – the second resonant angular frequency of the string oscillator ($\omega_2 = 2\pi f_2$, $f_1 = f_2/2$)
$\tau$ – mechanical relaxation time of the string oscillator
$Q_2$ – mechanical quality factor ($Q_2 = \omega_2 \tau / 2$)
$l$ – the length of the string (metres)
$\eta$ – mass per unit length (kg/m)
$k_0$ – differential mechanical-displacement-to-voltage transfer function (V/m)
$k_c$ – common mode rejection factor (dimensionless)
$k_s$ – total signal transfer function (1/m)
$G_p$ – feedback proportional gain (dimensionless)
$G_i$ – feedback integral gain (dimensionless)
$\tau_f$ – feedback time constant ($\tau_f >> 2\pi/\omega$)
$k_f$ – feedback transfer function (T/m)
$\Delta t_f$ – sampling interval in the digital path of the feedback loop
$\delta\omega$ – the drive frequency detune error ($\delta w = \omega - \omega_2$)
$e_n(t)$ – input voltage noise of the mechanical-displacement-to-voltage transducer

We also introduce the following in-phase and quadrature signals and noise components of the system under consideration

$$I_s(t) = k_s(\overline{X_2(t)Sin(\omega t)}), \quad Q_s(t) = k_s(\overline{X_2(t)Cos(\omega t)})$$

$$\alpha(t) = (\overline{N_2(t)Sin(\omega t)}), \quad \beta(t) = (\overline{N_2(t)Cos(\omega t)})$$

(4)

where the over-scored products indicate variables after the IF demodulation stage.

As for the thermal noise correlation function (see Eq.(3)), the time correlation function of the in-phase and quadrature noise components in Eq.(4) is as follows

$$\langle \alpha(t_1)\alpha(t_2) \rangle = \langle \beta(t_1)\beta(t_2) \rangle = \frac{4k_B T}{l\eta\tau}\delta(t_1 - t_2)$$

(5)

The $I_s$ and $Q_s$ dynamic variables are the standard representation of much slower varying processes in the feedback loop compared to the string vibration time scale, i.e. - $2\pi/\omega$.

Using Eqs.(4) and some simple mathematical manipulations (see Appendix A), one can replace Eq.(2) with the following ones

$$\frac{d}{dt}Q_s + \frac{1}{\tau}Q_s = -\delta\omega I_s + \frac{1}{2\omega\tau}\frac{d}{dt}I_s + \frac{k_s}{2\omega}\frac{l}{2\pi}\frac{i_s}{\eta}B_{yz}(t) - \frac{k_s}{2\omega}\beta(t)$$

(6)

$$\frac{d}{dt}I_s + \frac{1}{\tau}I_s = \delta\omega Q_s - \frac{1}{2\omega\tau}\frac{d}{dt}Q_s + \frac{k_s}{2\omega}\alpha(t)$$

(7)

The effective magnetic gradient in Eq.(6) is the sum of an external magnetic gradient and one created by the feedback loop

$$B_{yz}(t) = B_{yz}^{ext}(t) + B_{yz}^f(t - \tau_f) \cong B_{yz}^{ext}(t) + B_{yz}^f(t) - \tau_f \frac{d}{dt}B_{yz}^f(t)$$

$$B_{yz}^f(t) = -k_f\left[G_p\delta Q_s(t) + \frac{G_i}{\Delta t_f}\int_0^t dt'\delta Q_s(t')\right]$$

$$\delta Q_s(t) = Q_s(t) - Q_s^0 = Q_s(t) - \frac{1}{k_f}B_{yz}^0$$

(8)

In Eqs.(8), total time delay $\tau_f$ in the feedback loop has been taken into account. Also, $Q_s^0$ is a DSMG signal set point. It represents a constant signal level (effectively a number), which is maintained by the feedback loop servo when the whole system "string oscillator + feedback" is in equilibrium.

By combining Eq.(6), Eq.(7) and Eqs.(8), one can get the following DSGM closed loop operation equations

$$(1-\frac{\tau_f}{\tau}G_p)\frac{d}{dt}\delta Q_s + \frac{1}{\tau}(1+G_p - \frac{\tau_f}{\Delta t_f}G_i)\delta Q_s + \frac{G_i}{\tau \Delta t_f}\int_0^t dt' \delta Q_s = \quad (9)$$

$$= \frac{1}{\tau}\frac{\delta B_{yz}(t)}{k_f} - \frac{1}{\tau}\frac{2\pi}{l}\frac{\eta}{i_s}\frac{\beta(t)}{k_f}$$

$$\frac{d}{dt}I_s + \frac{1}{\tau}I_s = \delta\omega Q_s^0 + \frac{1}{\tau}\frac{2\pi}{l}\frac{\eta}{i_s}\frac{\alpha(t)}{k_f} \quad (10)$$

$$\delta B_{yz}(t) = B_{yz}^{ext} - k_f Q_s^0 = B_{yz}^{ext} - B_{yz}^0 \quad (11)$$

Here we have ignored some second order terms as we only consider a small departure from the equilibrium state, when $I_s \sim 0$ ($\delta\omega \sim 0$), and assuming that the string's mechanical quality factor is much larger than unity.

The output signal of DSMG after the digital low-pass filter is

$$SignalOut = \frac{1}{\tau_m}\int_{t-\tau_m}^{t}\left(\frac{G_i}{\Delta t_f}\int_0^{t'} dt'' \delta Q_s\right) dt' \quad (12)$$

For the purpose of analytical description of the signal processing, the digital output low-pass filter is represented by an additional integrator as per Fig.1 and Eq.(12), where $\tau_m$ is the integration time.

By substituting

$$\delta Q_s(t) = \tau \frac{d}{dt}Y(t) \quad (13)$$

where $Y(t)$ is a new dynamic variable, we obtain, finally, the complete set of the equations that govern the DSMG feedback closed loop operation including the fundamental thermal noise source

$$\frac{d^2}{dt^2}Y + \frac{2}{\tau_{eff}}\frac{d}{dt}Y + \omega_{eff}^2 Y = \quad (14)$$

$$= \frac{1}{k_f}\frac{1}{1-\frac{\tau_f}{\tau}G_p}\frac{1}{\tau^2}\left(\delta B_{yz}(t) - \frac{2\pi}{l}\frac{\eta}{i_s}\beta(t)\right)$$

where

$$\frac{2}{\tau_{eff}} = \frac{1}{\tau}\frac{1+G_p - \frac{\tau_f}{\Delta t_f}G_i}{1-\frac{\tau_f}{\tau}G_p} \quad (15)$$

$$\omega_{eff}^2 = \frac{1}{\tau \Delta t_f} \frac{G_I}{1 - \frac{\tau_f}{\tau} G_p} \tag{16}$$

$$SignalOut = G_I \frac{\tau}{\Delta t_f} \frac{1}{\tau_m} \int_{t-\tau_m}^{t} Y(t') dt' \tag{17}$$

and the following useful relation has been used

$$k_s = \frac{1}{k_f} \frac{2\pi}{l} \frac{2\omega}{\tau} \frac{\eta}{i_s} \tag{18}$$

The system described by Eq.(14) is a standard oscillatory system, with the distinction that its parameters are determined by the proportional and integral feedback components. Also, the relaxation time of the string, the feedback's sampling rate and time constant, and the measurement time play an important role in optimizing the performance of a DSMG. Effectively, there are 6 degrees of freedom, which determine the dynamic properties of the whole system "string oscillator + feedback". One can show that the system is unstable when $G_i < 0$ irrespective of $G_p$, $\tau$, $\tau_f$ and $\Delta t_f$. The proportional gain can be both positive and negative. A simple analysis shows that the only following possible boundaries for $G_p$ and $G_i$, when the system "string oscillator + feedback" is stable, are

<u>Case A:</u> $G_p \geq 0$, $G_i \geq 0$ (19)
$0 \leq G_p < \tau/\tau_f$,  $0 \leq G_i < (\Delta t_f / \tau_f)(1+G_p)$

<u>Case B:</u> $G_p \leq 0$, $G_i \geq 0$ (20)
$0 \leq |G_p| < 1$,  $0 \leq G_i < (\Delta t_f / \tau_f)(1 - |G_p|)$

We now consider the conditions at which critical damping is achieved

$$\omega_{eff}^2 = \frac{1}{\tau_{eff}^2} \tag{21}$$

By substituting Eq.(15) and Eq.(16) into Eq.(21), one can find the critical value of the integral gain as a function of the proportional gain within the stable operation boundaries determined by Eq.(19) and Eq.(20)

$$G_i^{critical}(G_p) = \frac{\Delta t_f}{\tau_f}\left(1 + \frac{2\tau}{\tau_f} + G_p\right)\left(1 - \sqrt{1 - \left(\frac{1+G_p}{1+\frac{2\tau}{\tau_f}+G_p}\right)^2}\right) \tag{22}$$

Fig.3 shows the whole region of the feedback loop stable operation divided by an over-damped zone and an under-damped zone. As seen from Eq.(22), there are two important parameters in the time domain, namely

$$A = \Delta t_f / \tau_f \tag{23}$$
$$B = \tau / \tau_f \tag{24}$$

The first characterizes the relation between the data processing time in the digital domain and the delay (inertia) in converting the data into differential force acting upon the string oscillator. The second one characterizes the relation between the inertia, in the time domain, of the string oscillator reacting to that differential force and the inertia of the feedback loop itself. They both play an important role in designing an optimized DSMG.

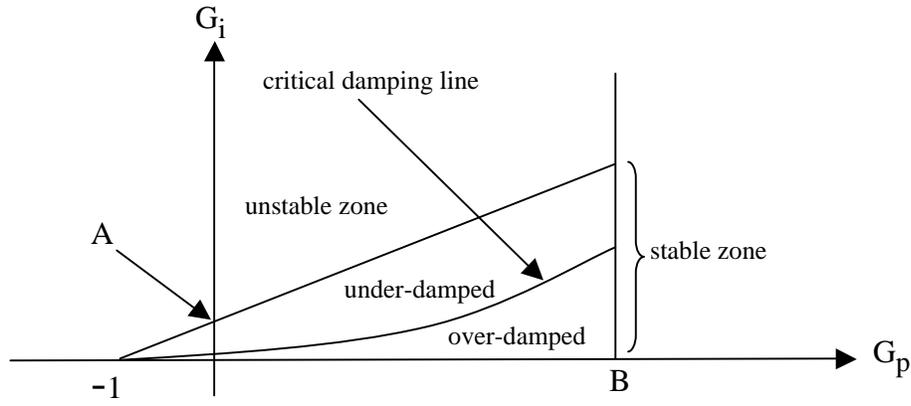

Fig. 3. The boundaries of the force feedback stable operation separated by an over-damped and an under-damped zone.

Both A and B are determined experimentally using the $G_i$ versus $G_p$ plot depicted in Fig.3. By setting the proportional gain to zero then gradually increasing the integral gain from the over-damped area to the under-damped area, the critical damping point can be detected when

$$G_i^{critical}(0) = A(1+2B)\left(1-\sqrt{1-\left(\frac{1}{1+2B}\right)^2}\right) \tag{25}$$

By increasing the integral gain further into the under-damped zone, the stable operation crossing point is reached at

$$G_i^{unstable}(0) = A \tag{26}$$

From Eq.(25) and Eq.(26), one can determine that

$$B = \frac{1}{4}\left(\frac{G_i^{unstable}(0)}{G_i^{critical}(0)}\right)\left(1-\frac{G_i^{critical}(0)}{G_i^{unstable}(0)}\right)^2 \tag{27}$$

Below we consider the fundamental thermal noise floor set by different combinations of the feedback components. It is important to point out that any feedback will introduce additional noise (back action noise). However, in a real situation, this noise is far below the level that is set by the practical requirements to the sensitivity of a DSMG. We will discuss this situation in a separate section at the end of this paper.

**3. Fundamental thermal noise floor of DSMG**

First, we consider the system "string oscillator + feedback" where both proportional and integral components of the feedback are set to zero. Instead of using Eq.(17) as an output signal, we shall use Eq.(13). When this "free" system is driven by noise only, Eq.(14) yields

$$\frac{d}{dt}\delta Q_s + \frac{1}{\tau}\delta Q_s = -\frac{1}{k_f}\frac{1}{\tau}\frac{2\pi}{l}\frac{\eta}{i_s}\beta(t) \tag{28}$$

The signal variance (in T/m) then can be calculated from Eq.(28) in the usual fashion [7]

$$\sigma_0^2 = k_f^2 \langle \delta Q_s^2 \rangle = \frac{1}{\tau^2}\frac{4\pi}{l^2}\frac{\eta^2}{i_s^2}\int_0^\infty d\Omega |F(\Omega)|^2 |Z(\Omega)|^2 \langle |\beta(\Omega)|^2 \rangle \tag{29}$$

$$|Z(\Omega)|^2 = \frac{1}{\Omega^2 + \frac{1}{\tau^2}} \tag{30}$$

$$|F(\Omega)|^2 = \frac{Sin^2(\Omega\tau_m/2)}{(\Omega\tau_m/2)^2} \tag{31}$$

where $Z(\Omega)$ – is the "free" system complex admittance, and $F(\Omega)$ – is the analytical representation of the digital low-pass output filter in the frequency domain.

The integration in Eq.(29) can be easily done by using the following tabulated integral [8]

$$\int_0^\infty dx \frac{Sin^2(x)}{x^2(x^2+a^2)} = \frac{\pi}{2}\frac{1}{a^2}\left(1 - \frac{1}{2a}\left(1 - e^{-2a}\right)\right), \quad Re\{a\} > 0 \tag{32}$$

This yields

$$\sigma_0 = \frac{4\pi}{i_s}\sqrt{\frac{\eta}{l^3}\frac{k_B T_0}{\tau\tau_m}} \tag{33}$$

where $T_0$ is the effective noise temperature

$$T_0 = T\left(1 - \frac{\tau}{\tau_m}\left(1 - e^{-\frac{\tau_m}{\tau}}\right)\right) \tag{34}$$

We now consider the case where the feedback proportional and integral components are not equal to zero. If the system is driven by noise only, we obtain the following signal variance

$$\sigma_f^2 = k_f^2 \left\langle SignalOut^2 \right\rangle = \tag{35}$$

$$= \frac{1}{\Delta t_f^2 \tau^2} \frac{4\pi \eta^2}{l^2} \frac{G_i^2}{i_s^2 \left(1 - \frac{\tau_f}{\tau} G_p\right)^2} \int_0^\infty d\Omega |F(\Omega)|^2 |Z(\Omega)|^2 \left\langle |\beta(\Omega)|^2 \right\rangle$$

where now

$$|Z(\Omega)|^2 = \frac{1}{(\Omega^2 - \omega_{eff}^2)^2 + \frac{4\Omega^2}{\tau_{eff}^2}} \tag{36}$$

and $F(\Omega)$ is the same as in Eq.(31).

The integration in Eq.(35) is more complicated. One can prove that the expression for $|Z(\Omega)|^2$ in Eq.(36) can be broken down into two terms where each of them can be integrated using the same tabulated integral as we used above

$$|Z(\Omega)|^2 = \frac{1}{4} \frac{\tau_{eff}}{\gamma} \left( \frac{1}{\Omega^2 + \frac{1}{\tau_{eff}^2} + \gamma^2 - \frac{2}{\tau_{eff}}\gamma} - \frac{1}{\Omega^2 + \frac{1}{\tau_{eff}^2} + \gamma^2 + \frac{2}{\tau_{eff}}\gamma} \right) \tag{37}$$

where

$$\gamma = \begin{cases} \sqrt{\frac{1}{\tau_{eff}^2} - \omega_{eff}^2}, & \text{in the overdamped zone} \\ 0, & \text{at critical damping} \\ i\sqrt{\omega_{eff}^2 - \frac{1}{\tau_{eff}^2}}, & \text{in the underdamped zone} \end{cases} \tag{38}$$

and $i=\sqrt{-1}$. The integration in Eq.(35) yields

$$\sigma_f(\gamma) = \frac{4\pi}{i_s}\sqrt{\frac{\eta}{l^3}\frac{k_B T_f(\gamma)}{\tau\tau_m}} \tag{39}$$

where the effective noise temperature $T_f$ is as follows

$$T_f(\gamma) = T\left(1 + \frac{\tau_{eff}}{\tau_m}M(\gamma)\right) \tag{40}$$

and

$$M(\gamma) = \frac{1}{4}\frac{\omega_{eff}^4}{\gamma}\left[\frac{e^{-\tau_m\sqrt{\frac{1}{\tau_{eff}^2}+\gamma^2-\frac{2}{\tau_{eff}}\gamma}}-1}{\left(\frac{1}{\tau_{eff}^2}+\gamma^2-\frac{2}{\tau_{eff}}\gamma\right)^{3/2}} - \frac{e^{-\tau_m\sqrt{\frac{1}{\tau_{eff}^2}+\gamma^2+\frac{2}{\tau_{eff}}\gamma}}-1}{\left(\frac{1}{\tau_{eff}^2}+\gamma^2+\frac{2}{\tau_{eff}}\gamma\right)^{3/2}}\right] \tag{41}$$

For the over-damped case we have

$$M_{overdamped}(\gamma) = \frac{1}{4}\frac{\omega_{eff}^4}{\gamma}\left[\frac{e^{-\tau_m\left(\frac{1}{\tau_{eff}}-\gamma\right)}-1}{\left(\frac{1}{\tau_{eff}}-\gamma\right)^3} - \frac{e^{-\tau_m\left(\frac{1}{\tau_{eff}}+\gamma\right)}-1}{\left(\frac{1}{\tau_{eff}}+\gamma\right)^3}\right] \tag{42}$$

where $\gamma$ is a function of the $G_i$ and $G_p$ force feedback parameters

$$\gamma = \frac{1}{2\tau}\cdot\frac{\sqrt{(1+G_p-\frac{\tau_f}{\Delta t_f}G_i)^2 - \frac{4\tau}{\Delta t_f}G_i(1-\frac{\tau_f}{\tau}G_p)}}{1-\frac{\tau_f}{\tau}G_p} \tag{43}$$

In the under-damped case, we need to analytically continue the expression for $M(\gamma)$ in Eq.(41) into the complex plane by replacing $\gamma \to i\gamma$. This yields

$$M_{underdamped}(\gamma) = -\frac{\omega_{eff}}{2}\frac{Sin(3 Arctg(\gamma\tau_{eff}))}{\gamma}\left[1 - \right.$$

$$\left. - \frac{Sin(\tau_m\omega_{eff} Sin(Arctg(\gamma\tau_{eff})) + 3 Arctg(\gamma\tau_{eff}))}{Sin(3 Arctg(\gamma\tau_{eff}))}e^{-\tau_m\omega_{eff} Cos(Arctg(\gamma\tau_{eff}))}\right] \tag{44}$$

where

$$\gamma = \frac{1}{2\tau} \frac{\sqrt{\frac{4\tau}{\Delta t_f} G_i (1 - \frac{\tau_f}{\tau} G_p) - (1 + G_p - \frac{\tau_f}{\Delta t_f} G_i)^2}}{1 - \frac{\tau_f}{\tau} G_p} \quad (45)$$

The critical damping case is a degenerated form of Eq.(44) at $\gamma \to 0$:

$$M(0) = \frac{3}{2}\left(\left(1 + \frac{1}{3}\frac{\tau_m}{\tau_{eff}}\right) e^{-\frac{\tau_m}{\tau_{eff}}} - 1\right) \quad (46)$$

## 4. Signal response of DSMG and electronic cooling effects

By its very nature, a DSMG is a modulation-demodulation device. Like fluxgate magnetic gradiometers [9], it is capable of detecting quasi-DC magnetic gradients both in relative and in absolute units. This is the area of interest in the frequency domain, where any magnetic gradiometer should exhibit very low $1/f$ and $1/f^2$ noise.

In the 0.0002–0.2 Hz frequency range (typical for most of the space-borne [10] and terrestrial [11] magnetic gradiometry applications) there are a number of different strategies for optimum detection of signals with different characteristics (unknown shape and period signals, signals with known period and unknown amplitude and phase, random time arrival signals with known amplitude, etc., see [12]). Below we consider a case that has been a default set up in the current DSMG signal processing. However, it may not be the optimum detection case for the particular type of signals covering the frequency range above.

Let us consider a particular region of the feedback parameters in which

$$\frac{\omega_{eff}}{2\pi} >> \frac{1}{Q_{eff}} f_{max}, \quad Q_{eff} = \frac{\omega_{eff} \tau_{eff}}{2} \quad (47)$$

where $f_{max}$ is the maximum signal frequency of interest.

The DSMG response is as follows

$$k_f Y(t) \cong \frac{1}{\tau^2} \frac{1}{1 - \frac{\tau_f}{\tau} G_p} \frac{1}{\omega_{eff}^2} \delta B_{yz}(t) \quad (48)$$

By using Eq.(16) and Eq.(17), we obtain

$$SignalOut(t) = \frac{1}{\tau \Delta t_f} \frac{G_i}{1 - \frac{\tau_f}{\tau} G_p} \frac{1}{\omega_{eff}^2} \frac{1}{\tau_m} \int_{t-\tau_m}^{t} \delta B_{yz}(t')dt' = \langle \delta B_{yz}(t) \rangle_{\tau_m} \quad (49)$$

In this case, the output signal does not depend on any feedback parameters, and is a direct magnetic gradient reading averaged over the integration time $\tau_m \sim 1/2f_{max}$.

Furthermore, Eq.(40), Eq.(42), Eq.(44) and Eq.(46) impose an absolute limit on the magnetic gradient resolution of a DSMG operating under the conditions discussed above. As is well known (see [6] and [13] for example), a dissipative system incorporated into a force feedback loop can have a noise temperature lower than the absolute equilibrium temperature, T. This "electronic cooling" effect results from the fact that only the whole system, like "string oscillator + feedback", is in thermodynamic equilibrium, while some parts of it can operate outside of the equilibrium state. More detailed description of such "non equilibrium" measurements is beyond the scope of this paper.

**5. Common mode rejection of DSMG**

As can be seen from Eq.(2), there should be the string's off-resonance response to the drive force, at the drive frequency $\omega$, coupled to both the uniform magnetic induction component $B_y$ and the magnetic gradient $B_{yz}$ (common mode displacements of the string). It is assumed that the string's fundamental mode frequency is exactly half of the string drive frequency $\omega$. This condition is well satisfied for stretched strings [5]. From Eq.(2) we have

$$X_1(t) = -\frac{16}{3\pi} \frac{i_s}{\eta} \frac{1}{\omega^2} \left( Sin(\omega t) + \frac{4}{3} \frac{1}{Q_2} Cos(\omega t) \right) \left( B_y + \frac{l}{2} B_{yz} \right) \quad (50)$$

The string's common mode displacements are detected by a mechanical-displacement-to-voltage-transducer with a common mode rejection factor $k_c$ ($0 \le k_c \le 1$), and are then demodulated as false $I_s$ and $Q_s$ in-phase and quadrature signal components

$$I_s^{false} = k_c k_s \overline{(X_1(t)Sin(\omega t))}, \quad Q_s^{false} = k_c k_s \overline{(X_1(t)Cos(\omega t))} \quad (51)$$

This leads to the following error terms in the right hand side of Eq.(6) and Eq.(7)

$$\frac{d}{dt} Q_s + \frac{1}{\tau} Q_s = -\delta \omega I_s + \frac{k_s}{2\omega} \frac{l}{2\pi} \frac{i_s}{\eta} B_{yz} - \frac{k_s}{2\omega} \beta(t) -$$

$$-\frac{k_s}{2\omega} \frac{l}{2\pi} \frac{i_s}{\eta} \frac{k_c}{l} \frac{32}{3} \left( \frac{\delta \omega}{\omega} + \frac{2}{3} \frac{1}{Q_2^2} \right) \left( B_y + \frac{l}{2} B_{yz} \right) \quad (52)$$

$$\frac{d}{dt}I_s + \frac{1}{\tau}I_s = \delta\omega I_s + \frac{k_s}{2\omega}\alpha(t) + \frac{k_s}{2\omega}\frac{i_s}{\eta}\frac{8}{3\pi}\frac{k_c}{Q_2}\left(B_y + \frac{l}{2}B_{yz}\right) \tag{53}$$

We also need to take into account the phase error $\delta\phi$ that is introduced at the signal demodulation stage. This rotates the $I_s$ and $Q_s$ vectors in the I-Q plane from their default position as follows

$$I_s^{eff} = I_s Cos(\delta\varphi) - Q_s Sin(\delta\varphi) , \qquad Q_s^{eff} = Q_s Cos(\delta\varphi) + I_s Sin(\delta\varphi) \tag{54}$$

By first multiplying Eq.(52) by $Cos(\delta\phi)$ and Eq.(53) by $Sin(\delta\phi)$, and then summing, we have the following effective replacement for Eq.(52)

$$\frac{d}{dt}Q_s^{eff} + \frac{1}{\tau}Q_s^{eff} = -\delta\omega I_s^{eff} +$$

$$+ \frac{k_s}{2\omega}\frac{l}{2\pi}\frac{i_s}{\eta}\left\{B_{yz}\left[\left(1 - k_c\frac{16}{3}\left(\frac{\delta\omega}{\omega} + \frac{2}{3}\frac{1}{Q_2^2}\right)\right)Cos(\delta\varphi) + \frac{8}{3}\frac{k_c}{Q_2}Sin(\delta\varphi)\right] +$$

$$+ k_c\frac{16}{3}\frac{B_y}{l}\left[\frac{1}{Q_2}Sin(\delta\varphi) - 2\left(\frac{\delta\omega}{\omega} + \frac{2}{3}\frac{1}{Q_2^2}\right)Cos(\delta\varphi)\right]\right\} +$$

$$+ \frac{k_s}{2\omega}\left(\alpha(t)Sin(\delta\varphi) - \beta(t)Cos(\delta\varphi)\right) \tag{55}$$

Similarly by multiplying Eq.(53) by $Cos(\delta\phi)$ and Eq.(52) by $Sin(\delta\phi)$, and then subtracting, we have the following effective replacement for Eq.(53)

$$\frac{d}{dt}I_s^{eff} + \frac{1}{\tau}I_s^{eff} = \delta\omega Q_s^{eff} +$$

$$+ \frac{k_s}{2\omega}\frac{l}{2\pi}\frac{i_s}{\eta}\left\{-B_{yz}\left[\left(1 - k_c\frac{16}{3}\left(\frac{\delta\omega}{\omega} + \frac{2}{3}\frac{1}{Q_2^2}\right)\right)Sin(\delta\varphi) - \frac{8}{3}\frac{k_c}{Q_2}Cos(\delta\varphi)\right] +$$

$$+ k_c\frac{16}{3}\frac{B_y}{l}\left[\frac{1}{Q_2}Cos(\delta\varphi) + 2\left(\frac{\delta\omega}{\omega} + \frac{2}{3}\frac{1}{Q_2^2}\right)Sin(\delta\varphi)\right]\right\} +$$

$$+ \frac{k_s}{2\omega}\left(\alpha(t)Cos(\delta\varphi) - \beta(t)Sin(\delta\varphi)\right) \tag{56}$$

Hereafter we consider only the optimum case when the whole system "string oscillator + feedback" is in its equilibrium state, and all error terms are infinitesimally small

$$I_s^{eff} \sim 0 \tag{57}$$

$$Cos(\delta\varphi) \sim 1, \quad Sin(\delta\varphi) \sim \delta\varphi$$

Also, Eq.(56) is an open loop equation with respect to the feedback loop servo, so we can treat the magnetic gradient term as the default equilibrium quantity

$$B_{yz} \cong B_{yz}^0 = k_f Q_s^0 \tag{58}$$

This, however, does not apply to Eq.(55), which is the closed loop equation with respect to the feedback loop servo, so Eq.(8) should be used instead, where now

$$\delta Q_s \to \delta Q_s^{eff} = Q_s^{eff} - Q_s^0 \sim 0 \tag{59}$$

By using Eq.(57) and Eq.(58), we finally obtain the following first order corrections to the output functions of DSMGs including the common mode motion and the demodulation phase error effects

$$\delta B_{yz}^{measured}(t) = \delta B_{yz}^{ext}(t) - \frac{64}{9} \frac{k_c}{Q_2^2} \frac{B_y}{l} \tag{60}$$

$$\frac{\delta\omega}{\omega} = \frac{1}{2}\frac{1}{Q_2}\left(\delta\varphi - \frac{8}{3}\frac{k_c}{Q_2}\right) - \frac{8}{3}\frac{k_c}{k_f}\frac{1}{Q_2^2}\frac{B_y}{l} \tag{61}$$

By measuring the variation $\delta\omega/\omega$ when a DSMG rotates around its vertical axis in the Earth's magnetic field in the absence of large local magnetic gradients, one can experimentally determine the common mode rejection factor $k_c$ provided by a read-out system. From Eq.(60) we can define the optimum common mode rejection ratio as follows

$$\frac{1}{K_c} = \frac{9}{32}\frac{Q_2^2}{k_c} \tag{62}$$

Even for typically moderate mechanical Q-factor (~ 200) and the common mode rejection factor (~ 0.01) the optimum common mode rejection ratio of a DSMG is in excess of $10^6$.

**6. Read-out transducer noise and back action (feedback) noise**

There are a number of mechanical-displacement-to-voltage transducers, which are well suited for detecting transverse mechanical displacements of a stretched string [15]. Without specifying any particular one, we consider the read-out (instrumental) noise contribution to the noise floor of the whole system "string oscillator + feedback". The instrumental noise is not intrinsically generated within the feedback loop, and therefore it can be treated as an external parasitic signal. It is dependent on the whole read-out instrumentation design but, in the optimum case, only the first amplification stage should be the dominant noise source.

Assuming that the equivalent input voltage noise of a mechanical-displacement-to-voltage transducer is $e_n$, one can find that it is added to the desired signal in Eq.(14) as follows

$$\delta B_{yz}(t) = B_{yz}^{ext} - k_f \frac{k_s}{k_0}\left(G_p e_n(t) + \frac{G_i}{\Delta t_f}\int_0^t dt' e_n(t')\right) - k_f I_s^0 \tag{63}$$

where $k_0$ is the differential mechanical-displacement-to-voltage transfer function.

This results in the following effective noise term in the right hand side of Eq.(14)

$$\beta_{eff}(t) = \beta(t) + \frac{1}{k_0}\frac{2\omega}{\tau}\left(G_p e_n(t) + \frac{G_i}{\Delta t_f}\int_0^t dt' e_n(t')\right) \tag{64}$$

The main impact of the back-action noise on the noise performance of DSMG is that it now contains a non-stationary term. It is assumed that the equivalent read-out input voltage noise $e_n$ is delta-correlated (Gaussian white noise). However, after integration it is no longer delta-correlated, but rather Gaussian coloured noise. The impact of this non-stationary component depends on the feedback integral gain $G_i$. One can choose the minimum frequency $f_{min}$ of interest above which the relative contribution of such non-stationary noise, within the measurement time interval $\tau_m$, is smaller than that of the proportional gain term in Eq.(64)

$$\left\langle e_n^2(0 \leq t \leq \tau_m)\right\rangle > \frac{1}{\Delta t_f^2}\left(\frac{G_i}{G_p}\right)^2\left\langle\left(\int_0^{\tau_m} dt\, e_n(t)\right)^2\right\rangle \tag{65}$$

The rest of such non-stationary noise can be filtered out below $f_{min}$. There are a number of different advanced strategies for detecting signals in coloured noise [12], although a detailed description of possible solutions to the problem is beyond the scope of this paper. In fact, the inequality in Eq.(65) determines the white-noise-red-noise corner frequency of a DSMG, which exists anyway due to a variety of non-stationary processes in the system under consideration.

There is also an external noise component originating from quantization noise in the digital path of the feedback loop, and also provided by the DAC and feedback buffer (including the output low pass filter, see Fig.2). One can show that their contribution to the total noise pool of DSMG is much smaller compared to the read-out instrumental noise as they are coupled to it by a low gain feedback stage [6].

**7. Absolute measurements versus relative measurements**

Because it is a self-referencing device, the DSMG is capable of simultaneously measuring absolute and relative magnetic gradient values. During closed loop operation, the feedback loop servo maintains a default constant value of the magnetic gradient in order to lock the internal clock frequency $\omega$ on to the second resonant mode (signal mode) of the string oscillator. In the presence of a local external magnetic gradient (say $B_{yz}^{(local)}$), the system compensates for its magnitude and then establishes the default equilibrium state (set point) corresponding to the default magnetic gradient

$$B_{yz}^0 = k_f Q_s^0$$

The system then remembers the number (say $a_0$) that was used in order to compensate the local magnetic gradient $B_{yz}^{(local)}$ through the feedback loop, and reach the default signal set point. The absolute reading of a DSMG is then as follows

$$B_{yz}^{absolute} = B_{yz}^{local} + \delta B_{yz} = k_f(Q_s^0 - a_0) + \delta B_{yz}$$

where $\delta B_{yz}$ is the relative reading of a DSMG (see Eq.(50)). The absolute reading however may contain a systematic error originating from so-called built-in-magnet effects. This creates a built-in non-uniform magnetic field in the vicinity of the string oscillator due to magnetic impurities in different parts of a DSMG sensor. Such impurities can also create a variable component of the intrinsically built-in magnetic field. This occurs when, for example, the sensor rotates in the Earth's magnetic field and the impurities change their induced magnetization directions. There is a stand-alone engineering problem in creating built-in-magnet free DSMGs, vastly resolved by reducing the number of metal parts in the sensor module and by specially treating critical sensor components, such as with an acid wash [14].

## 8. Conclusions

This paper provides a detailed theoretical analysis of a novel Direct String Magnetic Gradiometer (DSMG). It opens up the possibility of optimizing the instrument's behaviour, obtaining the best noise performance and creating custom designed sensors aimed at specific applications, such as fundamental research and air-borne or space-borne deployable instruments.

Effectively, a DSMG is a force gradiometer as it measures the Ampere force gradient created by an external magnetic gradient and an electric current pumped into its single sensitive element – a stretched metal wire (string).

The sensor intrinsic noise floor is limited by thermally activated random vibrations of the wire in the vicinity of its second resonant (violin) mode. The noise floor is inversely proportional to $l^{3/2}$, where $l$ is the length of the wire. When there are no significant limitations on the size of the sensor (a number of ground-based, marine-based and space-based applications have no such limitations), its sensitivity can be vastly increased by an extended length. This is in contrast with the traditional method of creating a magnetic gradiometer by positioning two individual vector magnetometers along an extended base line, which imposes stringent requirements on the accuracy of the relative alignment of the magnetometers.

In the frequency domain the DSMG acts as an up-converter with regard to DC magnetic gradients and, therefore, can exhibit a very low 1/f noise. It also possesses an intrinsically inherent and relatively high common mode rejection ratio without use of any external compensation technique.

**Acknowledgements**

I wish to thank Dr Wayne McRae and Mr Howard Golden of Gravitec Instruments (AU) Pty Ltd. for their support in creating this paper. I am also thankful to Mr David Greager and Dr Barry Marlow of Industrial Research (NZ) Ltd for many years spent together on the development of the first digital prototypes of DSMGs incorporating the integral and proportional feedback. I do appreciate the help and support provided by Prof. David Blair, Dr Li Ju and Mr Andrew Sunderland of the University of Western Australia in recent years for further research and development in relation to this exciting instrument. My special acknowledgements to Mr Simon Fraser, Mr Clive Hayley and Dr Neil Fraser for many years of providing financial, legal and technical support for Gravitec Instruments (NZ) Ltd. where it all began.

**Appendix A**

Here we consider the exact correspondence between the IF demodulation stage in the digital domain presented in Fig.4 below, and the basic equations (6) and (7) in Section 2 of this paper.

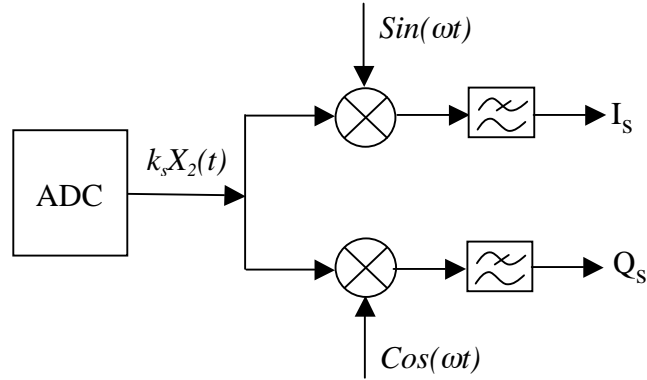

Fig.4. Schematic representation of the IF demodulation stage.

The input signal $k_s X_2(t)$ to the IF demodulation algorithm is related to the output in-phase and quadrature variables by the following convolution integrals

$$I_s(t) = k_s \int_{-\infty}^{\infty} dt' H(t-t') Sin(\omega t') X_2(t') \tag{A1}$$

$$Q_s(t) = k_s \int_{-\infty}^{\infty} dt' H(t-t') Cos(\omega t') X_2(t') \tag{A2}$$

where $H(t-t')$ is the impulse response a low pass filter as per Fig.4 (we assume the filters are linear and stationary, see also [12]).

If we multiply both left hand and right hand sides of Eq.2 by both Sin(ωt) and Cos(ωt), and then apply the convolution operators (A1) and (A2) to each term with the use of the following obvious equalities

$$Sin(\omega t)\frac{d^2}{dt^2} X_2(t) = \frac{d}{dt}\left(Sin(\omega t)\frac{d}{dt} X_2(t)\right) - \omega Cos(\omega t)\frac{d}{dt} X_2(t) \tag{A3}$$

$$Cos(\omega t)\frac{d^2}{dt^2} X_2(t) = \frac{d}{dt}\left(Cos(\omega t)\frac{d}{dt} X_2(t)\right) + \omega Sin(\omega t)\frac{d}{dt} X_2(t) \tag{A4}$$

$$Sin(\omega t)\frac{d}{dt} X_2(t) = \frac{d}{dt}\left(Sin(\omega t) X_2(t)\right) - \omega Cos(\omega t) X_2(t) \tag{A5}$$

$$Cos(\omega t)\frac{d}{dt} X_2(t) = \frac{d}{dt}\left(Cos(\omega t) X_2(t)\right) + \omega Sin(\omega t) X_2(t) \tag{A6}$$

we obtain

$$\frac{d}{dt}Q_s + \frac{1}{\tau}Q_s = \tag{A7}$$
$$= \frac{1}{2\omega}\frac{d}{dt}\left(\frac{d}{dt}I_s + \frac{1}{\tau}I_s\right) + \frac{1}{2\omega\tau}\frac{d}{dt}I_s - \frac{\omega^2 - \omega_2^2}{2\omega}I_s + \frac{k_s}{2\omega}\frac{l}{2\pi}\frac{i_s}{\eta}B_{yz}(t) - \frac{k_s}{2\omega}\beta(t)$$

$$\frac{d}{dt}I_s + \frac{1}{\tau}I_s = \tag{A8}$$
$$= -\frac{1}{2\omega}\frac{d}{dt}\left(\frac{d}{dt}Q_s + \frac{1}{\tau}Q_s\right) - \frac{1}{2\omega\tau}\frac{d}{dt}Q_s + \frac{\omega^2 - \omega_2^2}{2\omega}Q_s + \frac{k_s}{2\omega}\alpha(t)$$

Here, we have also used the fact that the convolution operators (A1) and (A2) and a time derivative operator of any order are commutative operators. Below we consider the case when the following inequalities are satisfied (which is the case for almost all range of the critical parameters that control optimum operation of a DSMG)

$$\frac{\delta\omega}{\omega} \equiv \frac{\omega - \omega_2}{\omega} \ll 1 \;,\; \frac{\pi}{\omega\Delta t_f} \ll 1 \;,\; \frac{1}{\omega\tau} \ll 1 \tag{A9}$$

and $1/2\Delta t_f$ is the Nyquist frequency of the data processing rate in the digital domain.

Finally, by using the following simplification

$$\omega^2 - \omega_2^2 \cong 2\omega\delta\omega \tag{A10}$$

and taking into account (A9), one can arrive at equations (6) and (7) as the starting point in the description of closed loop operation of a DSMG.